\documentclass[11pt,twoside]{article}
\usepackage{newpasp}
\usepackage{epsfig}

\index{Barnes, J. E.}

\begin{document}

\title{Dynamics of Stellar Collisions}

\author{J. E. Barnes}
\affil{Institute for Astronomy, University of Hawaii,
       2680 Woodlawn Drive, Honolulu HI, 96822}

\begin{abstract}
I compare gas-dynamical and stellar-dynamical models of collisions.
These two models have distinctly different physics; for example,
shocks introduce irreversibility in gas systems, while stellar systems
evolve in a completely reversible fashion.  Nonetheless, both models
yield broadly similar results, suggesting that analogies between gas
and stellar dynamics have some heuristic validity even applied to
collisions.
\end{abstract}

\keywords{stars:interactions -- hydrodynamics -- galaxies:interactions
-- stellar dynamics}

\section{Introduction}

Stars and galaxies are prototypical examples of self-gravitating
systems.  While both are held together by gravity, they differ in
size, makeup, and structure.  From a dynamical point of view, the
basic difference is that stars are made up of particles which {\it
often\/} undergo collisions, while galaxies are made up of particles
which almost {\it never\/} collide (this conference notwithstanding).

Analogies between stellar and gaseous systems run deep.  For example,
there is a close relationship between spherical stellar systems with
isotropic distribution functions $f = f(E)$ and gaseous systems with
barotropic equations of state $P = P(\rho)$.  The stability properties
of spherical stellar and gaseous systems provide further analogies; an
isotropic stellar system with $df/dE < 0$ is stable if the barotropic
gas-sphere with the same density profile is stable (Antonov 1962;
Lynden-Bell 1962).  This is a sufficient but not necessary condition;
on the whole, spherical stellar systems seem to be {\it more\/} stable
than their gaseous counterparts.  For flattened systems the situation
is more complex.  Infinite, uniformly rotating sheets of gas and stars
have very similar stability criteria (Toomre 1964).  On the other
hand, certain finite, uniformly-rotating stellar disks (Kalnajs 1972)
are much {\it less\/} stable than gaseous disks with the same density
profile.

\begin{figure}
\begin{center}
\epsfig{figure=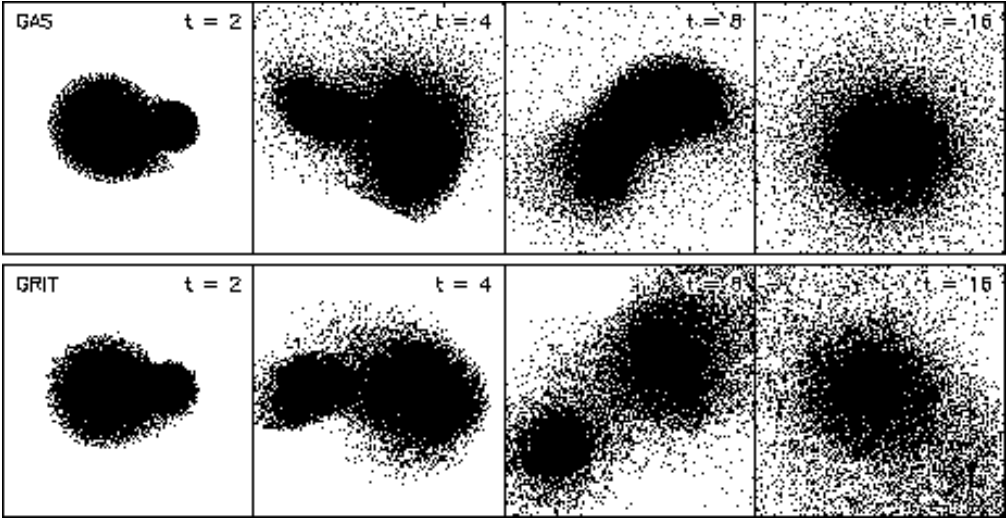,width=\textwidth}
\end{center}
\caption{\small Gas (top) and grit (bottom) versions of the same
collision.  Initially the two bodies approach on a parabolic relative
orbit, passing each other in a counter-clockwise direction.  The mass
ratio is $M_1/M_2 = 2$ and the targeted pericentric separation $r_{\rm
p} = 0.5$ is half the radius of the larger body.}
\label{fig01}
\end{figure}

Questions of existence and stability exploit analogies between stellar
and gaseous systems at or near equilibrium.  In this paper I use
numerical simulations to examine how such analogies work in situations
which are far from equilibrium: collisions.  Colliding stars and
colliding spherical galaxies evolve in superficially similar ways, as
seen in Fig.~\ref{fig01}.  Here the top row of frames shows a
collision of two gas-spheres, while the bottom row shows the analogous
collision of two spherical systems of collisionless particles,
hereafter labeled ``grit''\footnote{In my original presentation I used
``dust'' instead of ``grit'', borrowing this term from cosmology.
However, S.~Shapro pointed out that ``dust'' is only applicable to a
cold, pressureless medium.  Casting about for a suitable term, I chose
``grit'' since this conveys some resilience to pressure.}.  A more
detailed comparison of the two collisions in Fig.~\ref{fig01} reveals
significant differences.  For example, it's evident that orbit decay
doesn't work in quite the same way; contrasting the third pair of
frames, the two bodies are much closer in the gas model than they are
in the grit model.  Moreover, grit evolves {\it reversibly\/}, while
gas evolves {\it irreversibly\/}; by integrating the merged grit model
backward in time, one can unscramble the wreckage and recover the
initial conditions (van Albada \& van Gorkom 1977), but the presence
of shocks in the gas model precludes such reversals.

\subsection{Gas model}

Technically, the interior of a normal star is a partly-ionized plasma
in local equilibrium with a black-body radiation field; support
against gravity is provided by some combination of thermal and
radiation pressure.  But for low-mass stars the latter is
insignificant, and to a good approximation the pressure is given by
the equation of state for an ideal gas.

Gas dynamics unfolds in a space of three dimensions; the dynamical
variables are functions of position ${\bf r}$ and time $t$.  Two such
variables are the density $\rho = \rho({\bf r}, t)$ and the velocity
field ${\bf v} = {\bf v}({\bf r}, t)$.  One more variable is needed to
represent the thermodynamic state of the gas; in these calculations I
use the entropy function $a = a(S)$, which enters directly in the
equation of state:
\begin{equation}
  P = a(S) \rho^\gamma \, ,
  \label{equation-of-state}
\end{equation}
where $P$ is the pressure, and $\gamma$ is the ratio of specific
heats, here assigned the value $\gamma = 5/3$ appropriate for a
monatomic gas.

I adopt the following dynamical equations:
\begin{equation}
  \frac{\partial \rho}{\partial t} +
    \frac{\partial}{\partial {\bf r}} \cdot (\rho {\bf v}) = 0 \, ,
  \label{continuity-equation}
\end{equation}
\begin{equation}
  \frac{\partial {\bf v}}{\partial t} +
    \left( {\bf v} \cdot \frac{\partial}{\partial {\bf r}} \right) {\bf v} =
      - \frac{1}{\rho} \frac{\partial P}{\partial {\bf r}} -
        \frac{\partial \Phi}{\partial {\bf r}} \, ,
  \label{euler-equation}
\end{equation}
\begin{equation}
  \frac{\partial a}{\partial t} +
    \left( {\bf v} \cdot \frac{\partial}{\partial {\bf r}} \right) a =
      (\gamma - 1) \rho^{1 - \gamma} \, \dot{u}_{\rm v} \, .
  \label{entropy-equation}
\end{equation}
Here eq.~\ref{continuity-equation} represents conservation of mass and
eq.~\ref{euler-equation} represents conservation of momentum; the
gravitational potential $\Phi$ is calculated from Poisson's equation.
Eq.~\ref{entropy-equation} describes the evolution of the entropy
function; the flow is adiabatic, conserving the entropy of each fluid
element, except where the viscous heating function $\dot{u}_{\rm v} >
0$.  A standard form of artificial viscosity (Monaghan \& Gingold
1983) is used to implement the increase in gas entropy due to shocks.

I solve these dynamical equations using Smoothed Particle
Hydrodynamics or SPH (eg.~Monaghan 1992).  The code is similar to
``TREESPH'' (Hernquist \& Katz 1989); it includes a hierarchical
algorithm to compute gravitational forces, individual smoothing radii
$h_i$ set so that that each particle $i$ interacts with a fixed number
of neighbors, and individual time-steps adjusted to satisfy a Courant
condition.  The code doesn't include terms proportional to $\nabla
h_i$ which arise when the SPH equations are derived from a Hamiltonian
(Nelson \& Papaloizou 1993, 1994).  Since the code integrates
Eq.~\ref{entropy-equation} instead of the equivalent energy equation,
the neglect of these $\nabla h_i$ terms leads to imperfect energy
conservation (Hernquist 1993).  Most of the calculations presented
below conserve energy to $\sim 1$\% or better, so the neglect of these
terms is probably not critical in this application.

\subsection{Grit model}

The dominant mass components in typical galaxies are stars and dark
matter.  In the absence of conflicting evidence the latter is often
assumed to be composed of particles with masses much less than $10^6
M_\odot$ (cf.~Lacey \& Ostriker 1985).  The gravitational potential of
a galaxy is quite smooth and individual stars or dark matter particles
follow smooth trajectories, undisturbed by collisions, close
encounters, or any effect due to the discrete nature of the mass.

Stellar dynamics unfolds in a space of six dimensions; the phase-space
distribution $f = f({\bf r}, {\bf v}, t)$ is a function of position
${\bf r}$, velocity ${\bf v}$, and time $t$.  In the limit appropriate
for galaxies, $f$ evolves according to the Collisionless Boltzmann
Equation or CBE:
\begin{equation}
  \frac{\partial f}{\partial t} +
    {\bf v} \cdot \frac{\partial f}{\partial {\bf r}} -
      \frac{\partial \Phi}{\partial {\bf r}} \cdot
        \frac{\partial f}{\partial {\bf v}} = 0
  \label{vlasov-equation}
\end{equation}
Like eq.~\ref{continuity-equation}, eq.~\ref{vlasov-equation} is a
continuity equation, conserving the mass of the system.  But
eq.~\ref{vlasov-equation} describes an {\it incompressible\/} flow in
six dimensions; dynamics moves elements of phase-fluid around but
conserves the value of $f$ associated with each.

I solve the CBE using standard N-body techniques (eg.~Barnes 1998).
The code uses a hierarchical algorithm to compute gravitational forces
and a simple leap-frog integrator to advance particle coordinates.

\subsection{Polytropes}

I chose to collide simple polytropes, instead of using stellar models.
While accurate simulations of merging stars require realistic stellar
models (Sills \& Lombardi 1997), polytropes are still appropriate for
comparing the basic physics of stellar and galactic collisions.
Polytropes are easily constructed, and their properties are
well-understood (eg.~Chandrasekhar 1939).  Moreover, grit models of
polytropes have simple distribution functions (Eddington 1916) and the
stability of these systems has been studied using N-body simulations
(H\'enon 1973; Barnes, Goodman, \& Hut 1986).

Both gas and grit polytropes obey the following relationship between
mass density $\rho$ and gravitational potential $\Phi$:
\begin{equation}
  \rho = \rho(\Phi) = \rho_1 (1 - \Phi / \Phi_1)^n \, .
  \label{polytrope-density}
\end{equation}
Here $\Phi_1$ is the value of the potential on the surface of the
polytrope and $\rho_1$ is a constant with units of density.  By
construction, gas polytropes are marginally stable to convection, and
this implies that the index $n$ is related to the ratio of specific
heats by $n = 1 / (\gamma - 1)$.  For grit polytropes, $n$ is a free
parameter, and the distribution function takes the form
\begin{equation}
  f = f(E) =
    \left\{ \begin{array}{ll}
	     f_1 (1 - E / \Phi_1)^{n - 3/2} & \mbox{if $E<\Phi_1$,} \\
             0                              & \mbox{otherwise,}
           \end{array}
    \right.
  \label{polytrope-distribution}
\end{equation}
where $f_1$ is a constant with units of phase-space density.

For a monatomic gas, $\gamma = 5/3$, and the appropriate index $n =
3/2$.  The corresponding grit polytrope has an unusually simple
structure; $f$ has the constant value $f_1$ within the six-dimensional
volume defined by $E = \frac{1}{2}v^2 + \Phi(r) < \Phi_1$, and
vanishes everywhere else.

To construct realizations of polytropes I first tabulated the density
profile $\rho(r)$, potential $\Phi(r)$, and pressure $P(r)$.  Grit
models were generated by chosing initial positions ${\bf r}_i$ and
velocities ${\bf v}_i$ for each particle according to the distribution
function $f({\bf r}, {\bf v}) = f(\frac{1}{2}v^2 + \Phi(r))$ (H\'enon
1973).  Likewise, gas models were generated by chosing initial
positions ${\bf r}_i$ according to the density profile $\rho(r)$,
setting initial velocities ${\bf v}_i = 0$, and assigning initial
entropy function values $a_i = P(r_i) / \rho^\gamma(r_i)$.  In both
cases all particles had equal masses.  Since particle coordinates were
assigned independently, these initial configuration have Poissonian
fluctuations.  Such fluctuations are inevitable in N-body simulations,
but they are much larger than the fluctuations normally present in SPH
calculations.  Therefore, the gas realizations were relaxed by
evolving them with the SPH code using a velocity-damping term.  This
has the effect of ``ironing out'' the initial density fluctuations
without disturbing the overall density profile.

\section{Collision Sample}

The dynamics of a collision depend on the relative orbit and mass
ratio of the participants as well as their internal structures.  The
collisions reported here were restricted to parabolic (zero-energy)
orbits, which are appropriate for collisions in globular clusters and
similar environments.  I used $n = 3/2$ polytropes and assumed a
linear relationship between mass and radius; these choices roughly
caricature the properties of low-mass main-sequence stars.  It may
ultimately be worth relaxing these restrictions, but even with them in
place, two parameters must still be chosen for each collision: the
mass ratio and the pericentric separation of the initial relative
orbit.  I adopted mass ratios $M_1/M_2 = 1$ and~$2$, and studied
collisions with pericentric separations $r_{\rm p}$ ranging from zero
(head on) to $R_1 + R_2$ (grazing).  To simplify comparisons between
collisions with different mass ratios, I stipulated that all cases
have the same total mass.

Since $G$ is the only dimensional constant in these calculations, it's
convenient to adopt units with $G = 1$.  The polytropes used for the
equal-mass collisions ($M_1/M_2 = 1$) have masses $M_1 = M_2 = 0.75$
mass units and radii $R_1 = R_2 = 0.75$ length units; for the gas
models, the initial entropy function value is $a(S_1) = a(S_2) =
0.2889$.  The polytropes used for the unequal-mass collisions
($M_1/M_2 = 2$) have masses $M_1 = 1.0$ and $M_2 = 0.5$ and radii $R_1
= 1.0$ and $R_2 = 0.5$; the initial entropy function values are
$a(S_1) = 0.4240$ and $a(S_2) = 0.1682$.

For both mass ratios I considered four collisions with pericentric
separations $r_{\rm p} = 0$, $0.5$, $1$, and~$1.5$ ($= R_1 + R_2$);
each collision was run with both gas and grit models, using $24576$
particles for each calculation.  I ran all collisions with $r_{\rm p}
\le 1$ until the participants merged and relaxed to near-equilibrium
configurations.  The grazing collisions were not run to completion;
orbit decay is very gradual in such wide passages and the time
required to reach merger seemed excessive.

\begin{figure}
\begin{center}
\epsfig{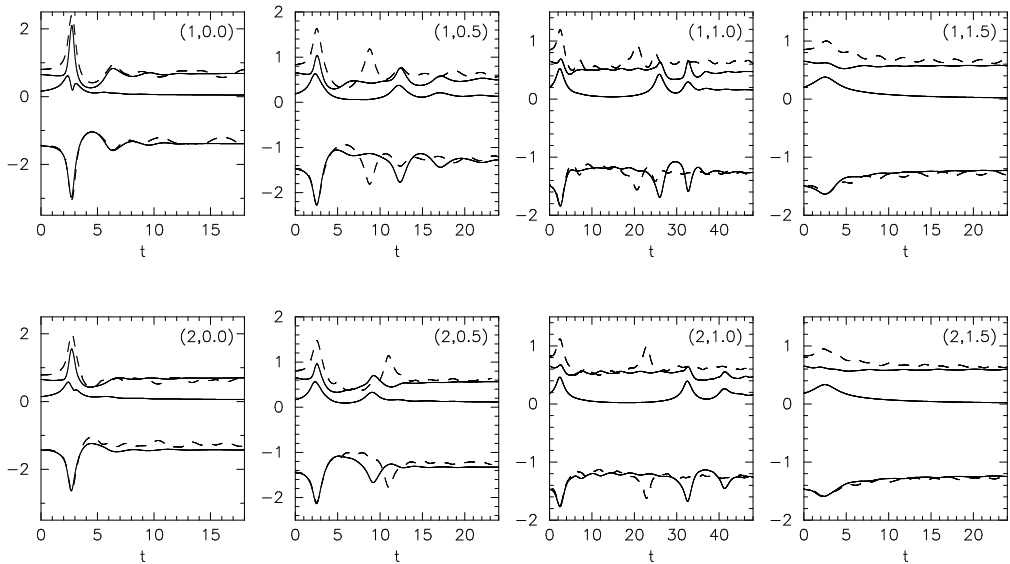}
\end{center}
\caption{\small Energies vs.~time for all collisions, each labeled by
$(M_1/M_2, r_{\rm p})$.  The three solid curves in each panel are from
the gas models; from top to bottom, they show the thermal energy
$T(t)$, kinetic energy $K(t)$, and potential energy $U(t)$.  Likewise,
the two dashed curves are from the grit models; they show the kinetic
and potential energies.}
\label{fig02}
\end{figure}

Fig.~\ref{fig02} summarizes the evolution of the collisions.  In each
case the participants have a first pericentric passage at $t \simeq
2.5$, marked by sharp minima in the curves of potential energies
$U(t)$.  In the grit models (dashed lines) this potential energy is
entirely invested in kinetic energy $K(t)$, which has a maxima
opposing the minima of $U(t)$.  In the gas models (solid lines) the
gravitational energy is shared between $K(t)$ and the thermal energy
$T(t)$, with the relative apportionment depending on the collision
parameters.  Head-on collisions produce sharp peaks in $T(t)$ as
gravitational energy heats the gas; the stored thermal energy drives a
subsequent re-expansion and bounce before the system settles down.  On
the other hand, the collisions with $r_{\rm p} \ge 1$ put most of the
available gravitational energy into bulk motion, producing modest
peaks in $K(t)$ while barely disturbing $T(t)$.

\begin{figure}[t!]
\begin{center}
\epsfig{figure=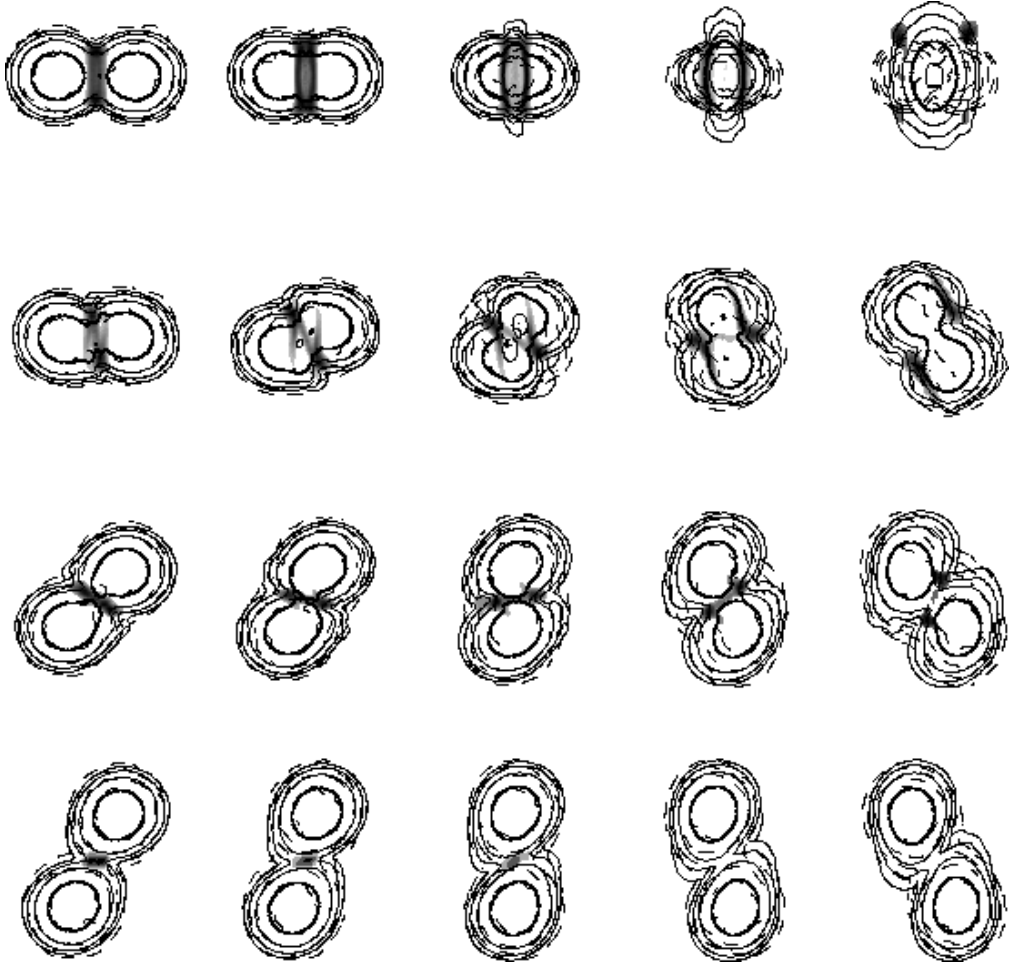,width=\textwidth}
\end{center}
\caption{\small First passages for equal-mass collisions.  Each row
shows a different collision at times equally spaced between $t = 2$
and $t = 3$; $r_{\rm p}$ increases from top to bottom.  Contours
indicate density on the orbital plane in steps of a factor of $4$;
dashed lines show grit, solid lines show gas, with a heavier contour
for $\rho = 1$.  Half-tones indicate shocks.}
\label{fig03}
\end{figure}

\begin{figure}[t!]
\begin{center}
\epsfig{figure=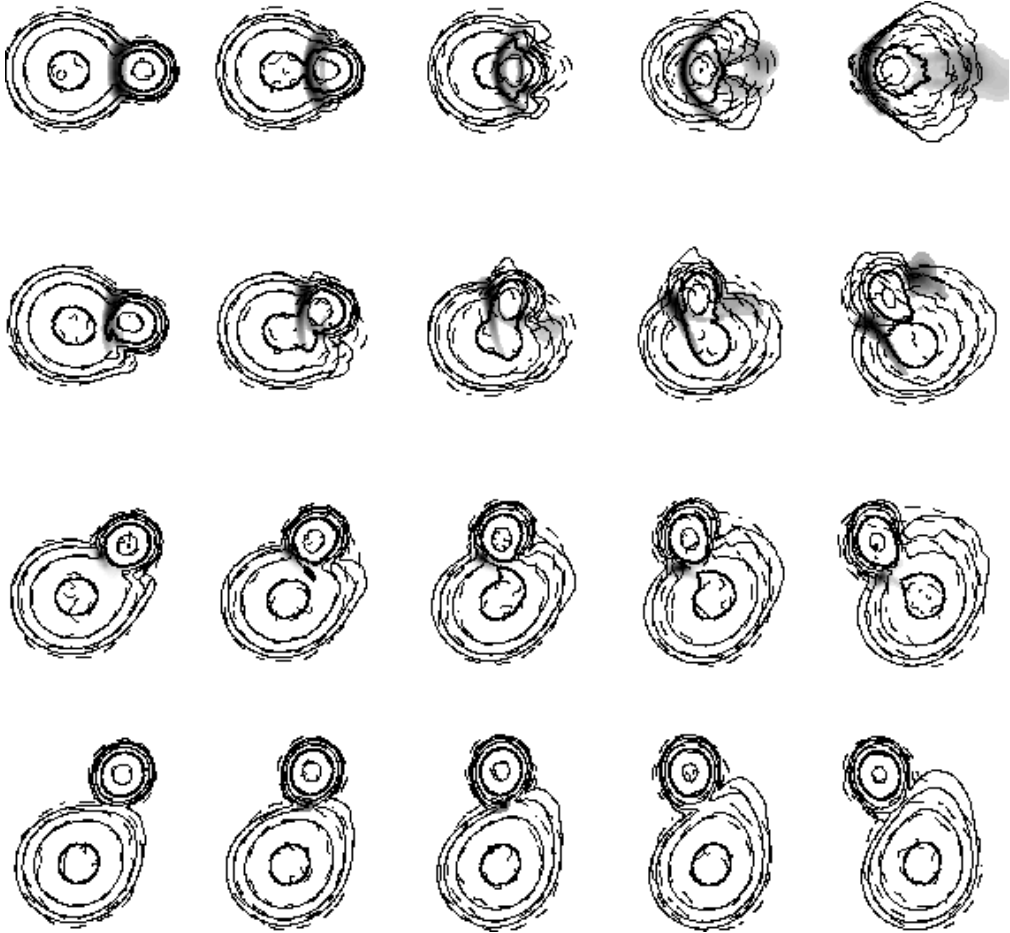,width=\textwidth}
\end{center}
\caption{\small First passages for unequal-mass collisions.  Each row
shows a different collision at times equally spaced between $t = 2$
and $t = 3$; $r_{\rm p}$ increases from top to bottom.  Contours
indicate density on the orbital plane in steps of a factor of $4$;
dashed lines show grit, solid lines show gas, with a heavier contour
for $\rho = 1$.  Half-tones indicate shocks.}
\label{fig04}
\end{figure}

\subsection{First passages}

Figs.~\ref{fig03} and~\ref{fig04} report on the first passages of all
collisions.  Here the contours represent density in the orbital plane,
evaluated using SPH-style interpolation; dashed contours show grit
density, while solid contours show gas density.  Shaded areas show
where $\dot{u}_{\rm v} > 0$; these track the shocks propagating
through the gas.

The most obvious differences between the gas and grit models occur in
the head-on collisions (top rows of Figs.~\ref{fig03}
and~\ref{fig04}).  In the equal-mass head-on collision the two grit
polytropes pass directly through each other.  As they do so they are
briefly compressed by their mutual gravity; at later times they
rebound, spraying grit particles onto loosely-bound orbits, and fall
back to form a single merged object (van Albada \& van Gorkom 1977).
The gas polytropes obviously can't interpenetrate; a shock develops
when they first touch.  Subsequently the shocked gas forms a disk
perpendicular to the collision axis and bounded on the left and right
by strong shocks.  In the final image of this collision (upper right
in Fig.~\ref{fig03}) no trace of the two gas polytropes remains; the
shocked gas forms a single object which expands due to its large
thermal energy content.  On the other hand, the grit polytropes,
having passed through each other, are still distinct.

The unequal-mass head-on collision produces a more complex morphology.
As the small grit polytrope plunges through its larger partner it
generates a ``wake'' -- a region of higher density where gravitational
focusing concentrates grit from the large polytrope towards the
collision axis.  This wake is similar to the mutual compression of the
two grit polytropes in the equal-mass collision; in both cases the
energy needed is extracted from orbital motion, resulting in rapid
orbit decay.  The gas model produces a very different result; as the
small polytrope plows into its partner it creates a strong bow shock.
This shock pushes gas {\it away\/} from the collision axis, evacuating
a cavity in the wake of the small gas polytrope.  Meanwhile, a weaker
shock travels backward through the small polytrope.  By the last image
(upper right in Fig.~\ref{fig04}) the bow shock has reached the
surface of the large polytrope, while the cavity has been filled in;
the small gas polytrope has lost most of its forward momentum and
lies, still largely intact, near center of the combined object.  In
contrast, at this time the grit polytropes have temporarily separated,
though they will soon fall back together.

Many of the dynamical effects just described also play a role in the
off-axis collisions.  In the equal-mass $r_{\rm p} = 0.5$ collision
the bodies of the two gas polytropes are traversed by parallel shocks
which slow their relative motion and increase their entropy.  The
unequal-mass version of this collision produces an asymmetric bow
shock as the small polytrope moves through the outer layers of its
companion, leaving a temporarily furrow behind.  In contrast, the grit
versions of these collisions again show regions of enhanced density in
the wakes of the interpenetrating polytropes.  Less dramatic effects
are seen in the wide collisions; the $r_{\rm p} = 1$ collisions are
basically ``kinder and gentler'' versions of their closer
counterparts, with weaker shocks and more subtle wakes.  Finally, the
grazing collisions ($r_{\rm p} = 1.5$), shown at the bottom of
Figs.~\ref{fig03} and~\ref{fig04}, don't develop significant shocks;
these passages involve only gravitational interactions.

\subsection{Orbit decay}

Although their initial orbits are parabolic, all of the off-center
collisions leave the participants on bound orbits after their first
passage.  This orbit decay is generally due to the transfer of orbital
energy to internal degrees of freedom, but the way this transfer takes
place may be quite different in gas and grit models.

From Fig.~\ref{fig02} it's already clear that the interplay between
gravity and gas-dynamics in orbit decay is complex.  The plots of
$U(t)$ for the collisions with $r_{\rm p} = 0.5$ and~$1.0$ show two or
more minima, each representing a different pericentric passage.  In
three of these collisions the grit models return to pericenter {\it
before\/} the corresponding gas models, but the {\it opposite\/} order
is seen in the $(M_1/M_2, r_{\rm p}) = (2,0.5)$ collision.  Thus
compared to the grit models, gas can either accelerate or delay orbit
decay.

\begin{figure}[t!]
\begin{center}
\epsfig{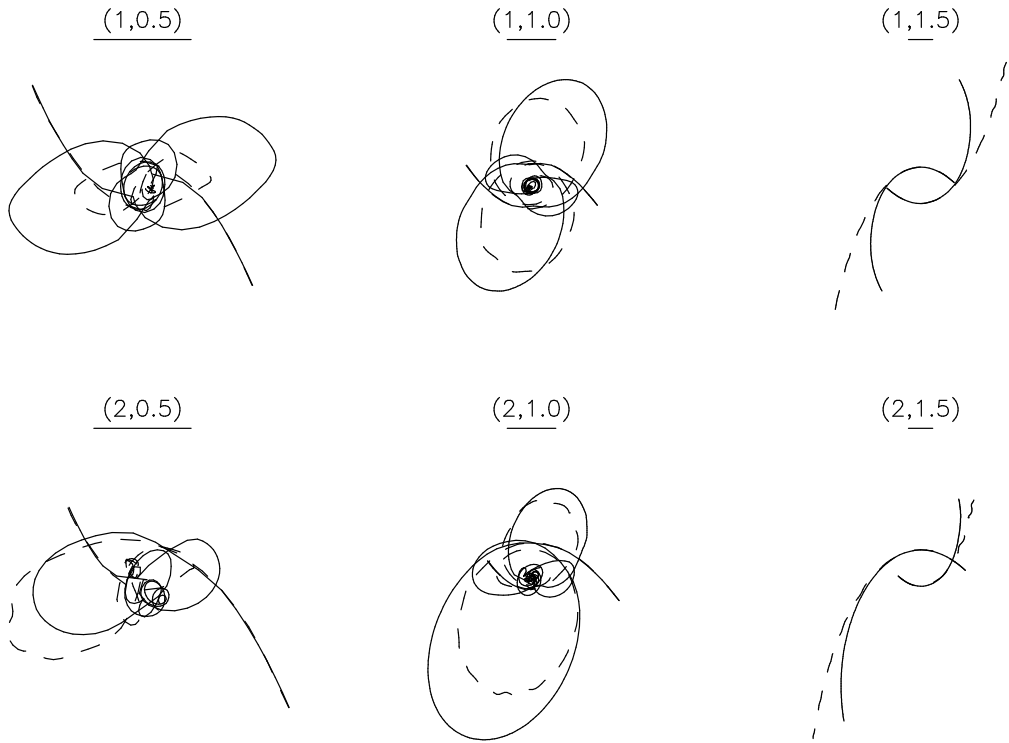}
\end{center}
\caption{\small Orbital trajectories for all off-center collisions.
Solid lines are gas results, while dashed lines are grit results.
Each is labeled by $(M_1/M_2,r_{\rm p})$; the scale bar under each
label is 1 unit long.}
\label{fig05}
\end{figure}

Fig.~\ref{fig05} shows orbital trajectories for all off-center
collisions.  These trajectories confirm that in the $(M_1/M_2,r_{\rm
p}) = (1,0.5)$, $(1,1.0)$, and $(2,1.0)$ collisions the grit models
undergo more rapid orbit decay than their gas counterparts, and that
the opposite is true for the $(2,0.5)$ collision.  Moreover, they
reveal a further puzzle; in grazing collisions the outgoing orbits of
the gas polytropes are more tightly bound than those of their grit
doppelgangers.

Closer inspection of Fig.~\ref{fig05} suggests a mechanism which might
delay orbit decay in some gas models.  Collision $(M_1/M_2,r_{\rm p})
= (1,0.5)$, in which the difference between the gas and grit models is
quite dramatic, shows it most clearly: on their first passage the gas
polytropes ``bounce'' off each other as if elastic.  Of course, this
is not too surprising; an ideal gas is {\it perfectly\/} elastic in
the absence of shocks.  Moreover, the corresponding panel of
Fig.~\ref{fig02} shows a sharp peak in the thermal energy $T(t)$
during this passage, and Fig.~\ref{fig03} shows that the gas is
compressed to roughly twice its initial density at the moment of
closest approach ($t \simeq 2.5$).  Similar bouncing trajectories are
seen in the other off-axis collisions.

On the other hand, the rapid orbit decay of the $(M_1/M_2,r_{\rm p}) =
(2,0.5)$ collision points to a different effect.  Here as in the
head-on cases the orbits of the gas polytropes decay faster than the
orbits of their grit counterparts.  For the head-on collisions this is
no great mystery; ram pressure brings the gas polytropes to a
screeching halt on their first and only passage.  A similar
explanation probably applies in this deeply-penetrating off-axis
collision; the small polytrope does so much work plowing through the
body of its large partner that it loses most of its orbital momentum.

The puzzle of the grazing collisions can't be explained in this way as
the gas polytropes barely touch each other.  These decays are governed
by tides, and it may seem paradoxical that the gas and grit models
with $r_{\rm p} = 1.5$ diverge since the same tidal gradients exist in
both cases.  However, the bottom rows of Figs.~\ref{fig03}
and~\ref{fig04} show that gas polytropes suffer stronger tidal
distortions in grazing collisions than do their grit counterparts.
This make sense; in the initial polytropes, gas particles are nearly
stationary, while grit particles are in constant motion.  Thus during
a tidal encounter, a well-placed gas particle can accumulate more
momentum than a wandering grit particle.  Now the stronger tidal
response of the gas polytropes is the key; recall that there would be
{\it no\/} orbit decay whatsoever if both participants remained
exactly spherical.  By being more responsive, gas polytropes couple
orbital motion to internal degrees of freedom more effectively, and
hence in grazing collisions their orbits decay faster than those of
grit polytropes.

\subsection{Later passages}

Thanks to orbit decay, later passages are slower and closer than first
ones.  Closer passages are always more violent, as Figs.~\ref{fig03}
and~\ref{fig04} amply illustrate.  On the other hand, the speed of
passage affects gas and grit models in different ways.  For grit
models, slower passages are generally {\it more\/} disruptive;
galaxies collide in exactly the way that cars don't, because low
speeds give tides more time to act.  Gas models, though also sensitive
to tides, suffer less from shocks in slow collisions; the coalescence
of two gas polytropes is more like a gentle swirling together than a
head-long plunge.

Every passage pumps orbital energy into internal degrees of freedom,
distending both gas and grit polytropes.  After each passages, gas
systems relax into convectively stable configurations; low-entropy
material remains centrally concentrated, while high-entropy material
forms an extended envelope.  Grit systems also acquire distended outer
envelopes; eq.~\ref{vlasov-equation} conserves phase-space density, so
in phase-space such an envelope is really a thin ``hyper-ribbon'',
tidally extracted from the main body, which still has the high
phase-space density of its source.  Nonetheless, this ribbon comes to
resemble a smooth envelope as phase-mixing winds it into a tight
spiral.

In basic outline, final encounters resemble mild versions of head-on
collisions; either both partners are subsumed into a single object, or
one partner burrows into the center of the other.

\section{Remnant Structure}

In the aftermath of a merger the wreckage undergoes a last episode of
dynamical relaxation as it evolves towards an equilibrium
configuration.  This final convulsion generally begins at the center
and travels outward.  In a grit system potential fluctuations during
the final merger scatter particles onto loosely-bound radial orbits;
as these outgoing particles reach apocenter they create a caustic or
``shell''.  In a gas system the final merger may create a weak
outgoing shock, but much of the readjustment takes place at subsonic
velocities, with no attendant increase in entropy.

\subsection{Density profiles}

\begin{figure}[t!]
\begin{center}
\epsfig{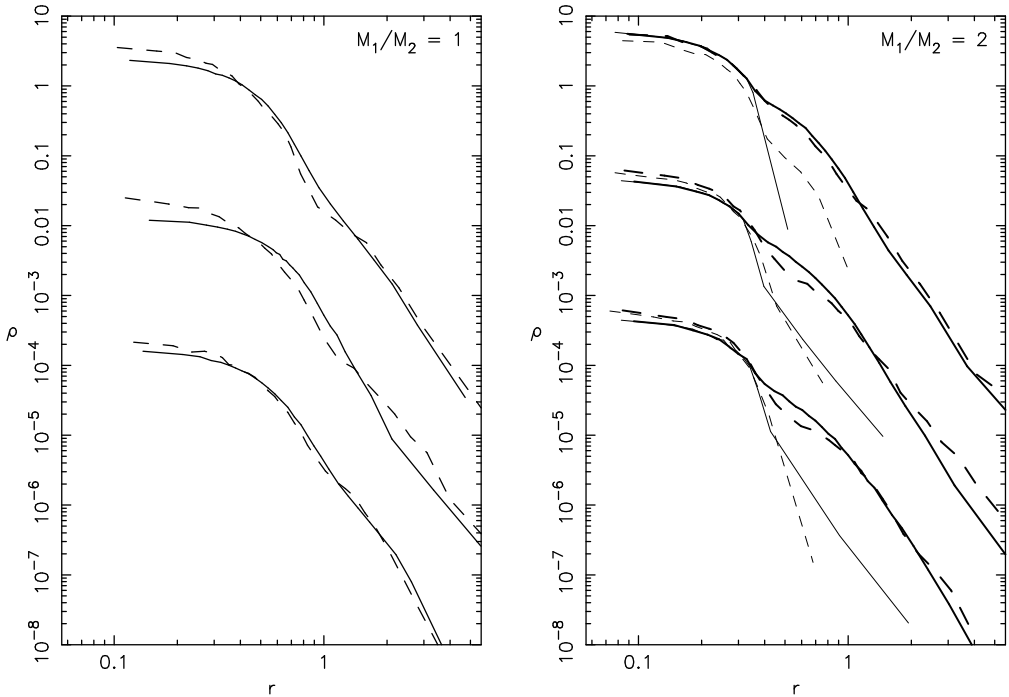}
\end{center}
\caption{\small Density profiles derived by spherical averaging.
Solid lines are gas results, while dashed lines are grit results.
Left: remnants of equal-mass collisions.  Right: remnants of
unequal-mass collisions; thin lines show profiles for particles from
the small polytropes.  Top to bottom: remnants of collisions with
$r_{\rm p} = 0$, $0.5$, and~$1$, displaced downward by successive
factors of $100$ for clarity.}
\label{fig06}
\end{figure}

Fig.~\ref{fig06} presents density profiles for all six merging
collisions (those with $r_{\rm p} \le 1$).  These profiles were
calculated by evaluating the mass in nested spherical shells.  Despite
the differing dynamics of gas and grit mergers, the resulting remnants
have similar profiles over a range of five decades in density.  At
large radii both grit and gas densities fall off roughly as $\rho
\propto r^{-4}$.  For grit models this asymptotic slope results as
phase-mixing spreads out a population of particles with a wide range
of energies including both bound and unbound orbits (Jaffe 1987; White
1987).

The unequal-mass mergers yield density profiles with ``shoulders''
where the slope briefly becomes shallower as $r$ increases.  These
compound profiles arise because the small polytropes resist disruption
during the merger process.  The thin lines in the right-hand panel of
Fig.~\ref{fig06} show density profiles derived using {\it only\/}
particles originating in from the small polytropes.  In every case,
the small polytropes have settled, virtually intact, in the centers of
the merger remnants.  For the gas models, this outcome is explained by
the lower entropy which the small polytrope retains throughout the
collision and merger; the remnant can't be convectively stable unless
this low-entropy material winds up at the center.

\begin{figure}
\begin{center}
\epsfig{figure=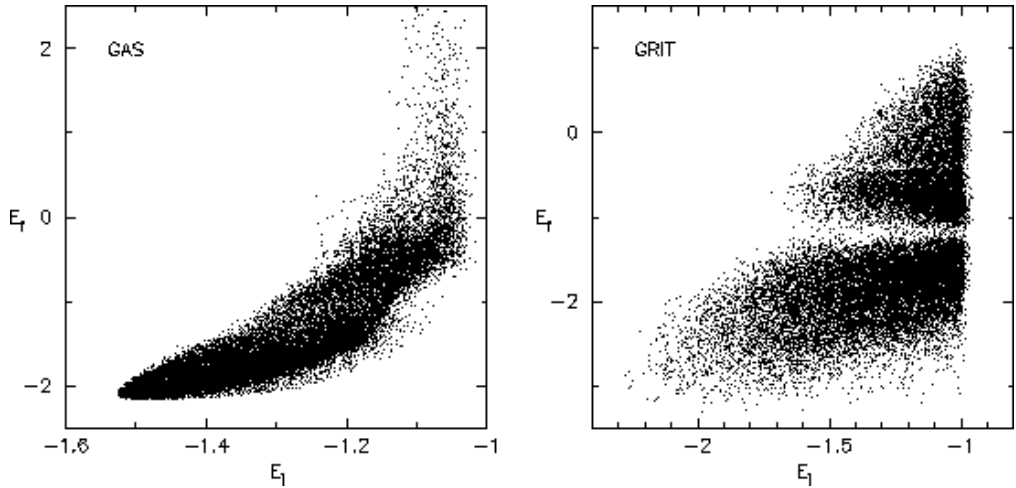,width=\textwidth}
\end{center}
\caption{\small Initial vs.~final binding energies for particles from
remnants of $(M_1/M_2,r_{\rm p}) = (1,0.0)$ collision.  Left: gas
model; right: grit model.  Note that the two plots have different
scales.  In both plots, $E = 0$ is escape energy.}
\label{fig07}
\end{figure}

If both partners have the {\it same\/} entropy, how much rearrangement
takes place during a merger?  Fig.~\ref{fig07} presents scatter-plots
of initial binding energy $E_{\rm i}$ vs.~final binding energy $E_{\rm
f}$ for particles in both versions of the $(M_1/M_2,r_{\rm p}) =
(1,0.0)$ collision.  Here $E_{\rm i}$ was measured with the polytropes
at infinite separation, while $E_{\rm f}$ was measured from the merger
remnant; internal energies were included in calculating binding
energies of gas particles.  These plots show that initial binding
energy is a useful, albeit imperfect, predictor of final binding
energy.  For the gas model, the correlation between $E_{\rm i}$ and
$E_{\rm f}$ is fairly strong; moreover, there is a one-to-one
relationship between binding energy and radius, so the initial and
final radii of gas particles are also correlated.

For the grit model things are more complex.  As the right-hand panel
of Fig.~\ref{fig07} shows, grit particles have a {\it bimodal\/}
distribution of final binding energies.  This arises because the
orbital phases of individual grit particles have a strong bearing on
their response to tidal interactions.  For example, consider a
particle which reaches an apocenter of its orbit just before the
polytropes interpenetrate, and the subsequent pericenter just after
they separate.  Such a particle sees a deeper potential well while
falling in than it does while climbing out, so it gains energy;
conversely, a particle on a similar orbit with a different phase may
lose energy.  The resulting bimodal energy distribution explains the
shoulder in the density profile of this remnant (Fig.~\ref{fig06},
left-hand panel, upper dashed curve).  Tightly-bound particles form
the core, while the rest form the $\rho \propto r^{-4}$ envelope; the
deficit of particles with $E_{\rm f} \simeq -1.2$ accounts for the dip
in $\rho(r)$ at $r \simeq 0.8$.

\subsection{Shapes}

\begin{figure}[t!]
\begin{center}
\epsfig{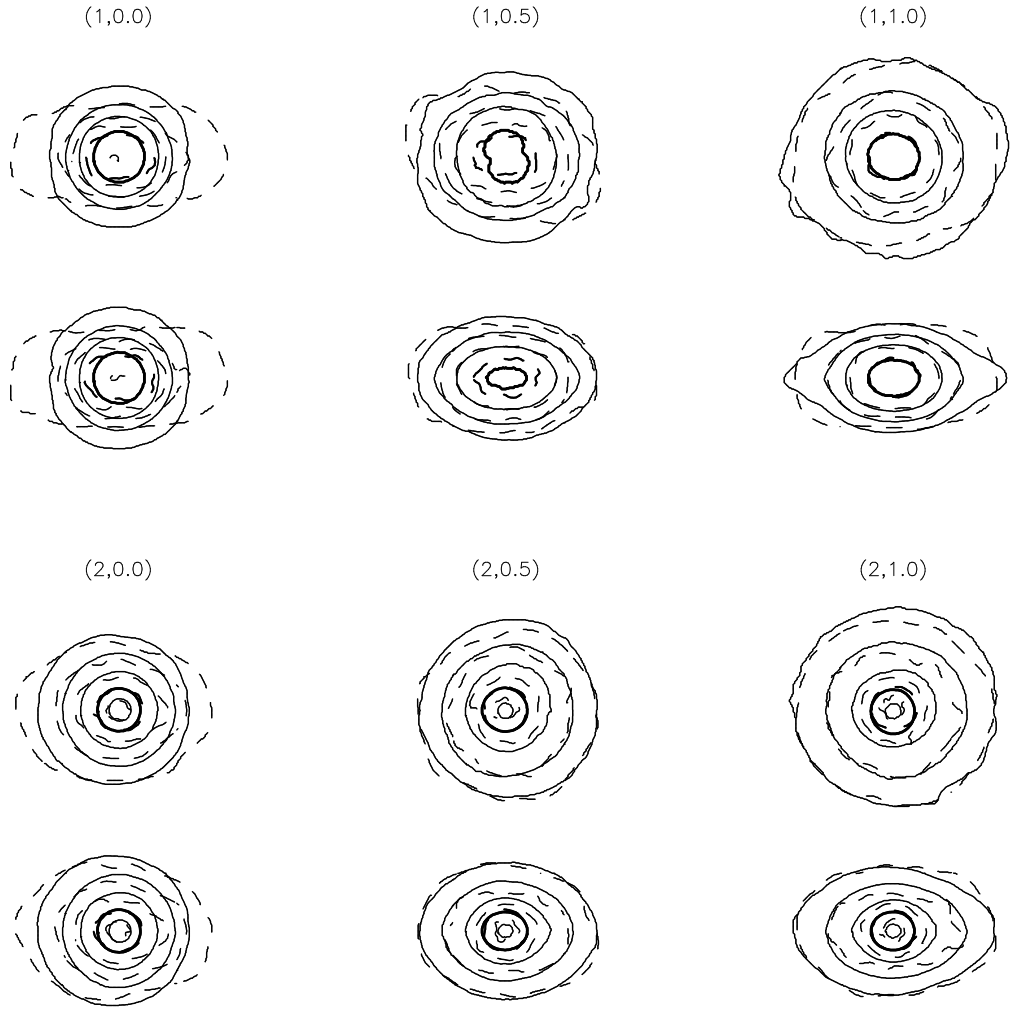}
\end{center}
\caption{\small Merger remnants.  Contours indicate densities in steps
of a factor of $4$; solid lines show gas, while dashed lines show
grit.  Each remnant is shown twice, once on a slice through the
orbital plane (above), and once on a perpendicular slice (below).}
\label{fig08}
\end{figure}

Fig.~\ref{fig08} presents density contours for the full sample of gas
and grit merger remnants.  Two slices through each remnant are shown;
one slice in the orbital plane, and one slice at right angles to this
plane.  As in Figs.~\ref{fig03} and~\ref{fig04}, here too the head-on
collisions produce the most striking discrepancies between gas and
grit models.  The head-on gas remnants are nearly spherical, as
expected for a self-gravitating gas configuration with no angular
momentum.  The head-on grit remnants, on the other hand, are prolate
bars aligned with the collision axis.  This is a ``long-term'' memory
of the original collision velocities.  Without the short-range
interactions which randomize gas motions, grit systems maintain
anisotropic velocity dispersions which can support such aspherical
shapes.

The remnants produced in off-axis collisions have significant amounts
of angular momentum.  Most such collisions yield fairly oblate
remnants; sliced perpendicular to the orbital plane, the gas remnants
are sometimes ``disky'' while the grit remnants seem more ``boxy'',
but in general the density contours are very similar.  The remnants
formed by the $(M_1/M_2,r_{\rm p}) = (1,0.5)$ collisions are somewhat
unusual in this regard.  Both are distinctly non-axisymmetric; sliced
along the orbital plane, the grit model has elliptical density
contours at all radii, while the gas model seems to have a doubled
core and a slightly elongated envelope.  The grit remnant may be
stable, since anisotropy and rotation together can support a tumbling
triaxial structure in equilibrium.  The same may not be true of the
non-axisymmetric gas remnant; more experiments could help settle this
issue.

\subsection{Kinematics}

\begin{figure}
\begin{center}
\epsfig{figure=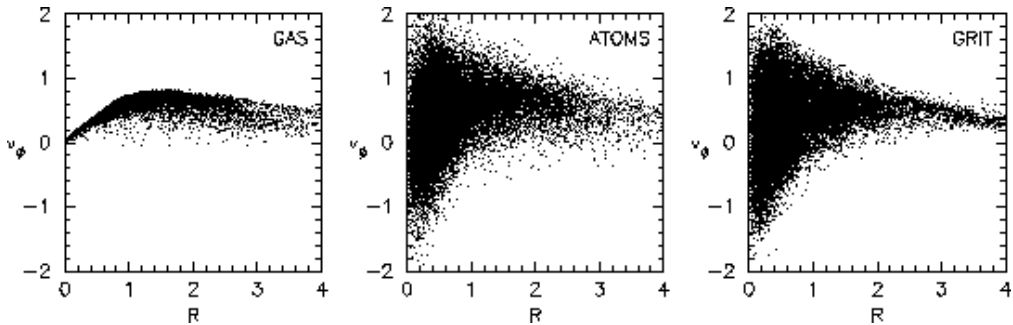,width=\textwidth}
\end{center}
\caption{\small Rotation velocities $v_\phi$ vs.~cylindrical radii $R$
for remnants of $(M_1/M_2,r_{\rm p}) = (1,1.0)$ collision.  Left: gas
model; middle: gas atoms; right: grit model.}
\label{fig09}
\end{figure}

Fig.~\ref{fig09} compares the kinematics of gas and grit versions of
remnants from collision $(M_1/M_2, r_{\rm p}) = (1, 1.0)$.  This
remnant has a relatively large amount of angular momentum and a
major-to-minor axis ratio of roughly $2$:$1$.  Here the velocity in
the direction of rotation, $v_\phi$, has been plotted against
cylindrical radius in the rotation plane, $R$.  Between the plots of
particle velocities for gas (left) and grit (right), an additional
plot shows velocities of gas atoms taking thermal motion into account
(middle).  The gas has a very regular velocity field, as reported in
previous studies (eg.~Lombardi, Rasio, \& Shapiro 1996).  In contrast,
the grit has a broad distribution of velocities at each point; while
the overall sense of rotation is the same as in the gas case, some
grit particles even counter-rotate.  But when the grit is compared to
the gas {\it atoms\/}, it's evident that these velocity distributions
are very much alike.  The similarity of these distributions is
consistent with the good match between the density contours of the gas
and grit versions of this remnant (Fig.~\ref{fig08}, upper-right).

\begin{figure}[t!]
\begin{center}
\epsfig{figure=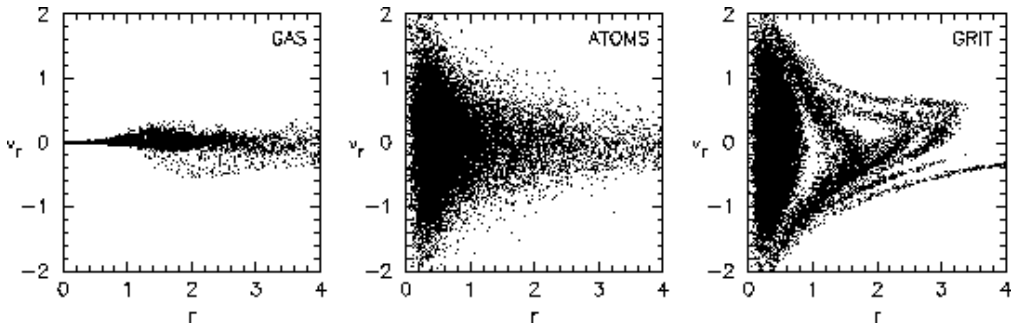,width=\textwidth}
\end{center}
\caption{\small Radial velocities $v_{\rm r}$ vs.~radii $r$ for
remnants of $(M_1/M_2,r_{\rm p}) = (1,0.0)$ collision.  Left: gas
model; middle: gas atoms; right: grit model.}
\label{fig10}
\end{figure}

Kinematics for the remnants of the head-on collision $(M_1/M_2, r_{\rm
p}) = (1, 0.0)$ are presented in Fig.~\ref{fig10}.  Here the radial
velocity, $v_{\rm r}$, has been plotted against the spherical radius,
$r$.  This remnant has no angular momentum, so gas velocities should
vanish; the small velocities seen in the left-hand panel indicate that
the remnant is not yet fully relaxed.  The grit version also is not
yet completely relaxed, since in perfect equilibrium the distribution
should be symmetric with respect to the line $v_{\rm r} = 0$.
Moreover, the grit plot shows that in phase-space the envelope of this
remnant is not smooth, but consists of narrow ribbons of particles.
These particles populate the upper peak in the final binding energy
distribution for this remnant (Fig.~\ref{fig07}, right-hand panel);
launched as the two polytropes interpenetrated, they fell back in
coherent streams and have since been wound up by ongoing phase-mixing.
With time this fine structure will become harder to see, and
eventually this plot will look something like the plot of atomic
velocities shown in the middle.  But even in the limit $t \to \infty$
some differences will remain; these include the radial anisotropy of
the grit velocity distribution and bimodal nature of its binding
energy distribution.

\section{Conclusions}

Comparison of gas-dynamical and stellar-dynamical models of colliding
systems reveals many specific differences and hints at some underlying
similarities.  While stellar systems interact via gravity alone, gas
systems interact via a complex mixture of gravity, pressure forces,
and shocks.  This mix greatly increases the range of dynamical
behavior in stellar collisions.  Even when interactions between gas
systems are limited to tides, the resulting deformations are stronger
than those seen in equivalent encounters of stellar systems.

Remnants of gas-dynamical and stellar-dynamical mergers also exhibit
significant differences.  Stellar systems have anisotropic velocity
distributions, supporting a wide range of remnant morphologies, while
gas systems settle into a small range of equilibria.  But at the same
time, the stellar and gaseous remnants studied here have very similar
density profiles, and often similar shapes as well.  This seems
remarkable, since the physics behind these density profiles is rather
different.  The extended envelopes of gaseous remnants contain
high-entropy material produced by shocks, which are violent,
well-localized events resulting from physical encounters.  On the
other hand, the envelopes of stellar remnants contain material
extracted from the participants by tides; this material retains the
high fine-grain phase-space density it starts with, and only when
coarse-grained does it approximate the atomic velocity distribution of
a gaseous envelope.  Two puzzles remain: first, why do such disparate
mechanisms -- shocks and tides -- produce such similar remnant
envelopes?  And second, how is coarse-graining -- an operation applied
to a stellar system by an observer -- analogous to the irreversible
physical processes which occur in gaseous systems?

\acknowledgements
I thank Piet Hut, James Lombardi, Frederic Rasio, and Stuart Shapiro
for helpful suggestions.  Support for this work was provided by NASA
grant NAG 5-8393 and by Space Telescope Science Institute grant
GO-06430.03-95A.

\end{document}